\documentstyle[preprint,aps,amsmath,amssymb,mathrsfs]{revtex}

\begin{document}
\preprint{KVI-1438/THEF-NYM 99.01}
\draft
\def\bfsig{\boldsymbol{\sigma}}
\def\bfr{\widehat{\boldsymbol{r}}}
\hyphenation{Nij-me-gen}
\hyphenation{Rij-ken}

\title{Chiral two-pion exchange and
       proton-proton partial-wave analysis}

\author{M.C.M. Rentmeester,$^{a,b,c}$ R.G.E. Timmermans,$^{a,d}$
        J.L. Friar,$^e$ J.J. de Swart$^b$}
\address{$^a$KVI, University of Groningen, Zernikelaan 25,
         \\ 9747 AA Groningen, The Netherlands}
\address{$^b$Institute for Theoretical Physics,
         University of Nijmegen,
         P.O. Box 9010, \\ 6500 GL Nijmegen, The Netherlands}
\address{$^c$Department of Physics, The Flinders University
         of South Australia, \\ Bedford Park, SA 5042 Australia}
\address{$^d$Institute for Theoretical Physics,
         University of Utrecht,
         Princetonplein 5, \\ 3584 CC Utrecht, The Netherlands}
\address{$^e$Theoretical Division, Los Alamos National Laboratory,
         \\ Los Alamos, NM 87545, U.S.A.}

\date{\today}
\maketitle
\begin{abstract}
The chiral two-pion exchange component of the long-range $pp$
interaction is studied in an energy-dependent partial-wave
analysis. We demonstrate its presence and importance, and
determine the chiral parameters $c_i$ ($i=1,3,4$). The values
agree well with those obtained from pion-nucleon amplitudes.
\end{abstract}

The longest-range part of the strong nucleon-nucleon ($N\!N$)
interaction is the well-established one-pion exchange (OPE)
force~\cite{Czi59,Tim93}. Next in range is the two-pion exchange
(TPE) force, the formulation of which has been a long-standing
problem~\cite{Mor61}, both in field theory~\cite{Cha59,Rij91}
and in dispersion theory~\cite{Ama60}.
In recent years, it has been argued that the key to the solution
is the chiral symmetry of QCD~\cite{Wei96}, and that the
long-range parts of the TPE potential can be derived
model-independently by a systematic expansion of the
effective chiral Lagrangian~\cite{Ord92}. In this
Letter, we will study this long-range chiral TPE force in
the proton-proton ($pp$) interaction and show unambiguously
its presence and its importance.

In the energy-dependent Nijmegen partial-wave analyses (PWA's)
of the $N\!N$ and $\overline{N}\!N$ scattering
data~\cite{Ber88,Ber90,Sto93,Tim94}, the long-range forces
are taken into account exactly and the short-range forces are
parametrized analytically. The partial-wave scattering amplitudes
are analytic functions of the energy. The nearby left-hand
singularities in the complex-energy plane are due to the
long-range forces; these cause the rapid energy dependence
of the physical $N\!N$ scattering amplitudes.
The shorter-range forces are responsible for the far-away
singularities, which give in the physical region only slow
energy variations of the amplitudes. This method of PWA can serve
as a sensitive tool to investigate precisely these long-range
interactions. It has been used successfully in studies of
electromagnetic interactions~\cite{Aus83} and of
the OPE potential~\cite{Tim93,Ber87,Swa97,Kol98}.
Here this tool will again be employed, now to study
the long-range chiral TPE component of the $pp$ force.

The methods of the Nijmegen PWA's are described in detail in
Refs.~\cite{Ber88,Ber90,Sto93,Tim94}. The long-range potentials,
including the full electromagnetic interaction (relativistic
Coulomb, magnetic-moment interaction, and vacuum polarization)
and the longest-range strong interactions
are used in the relativistic Schr\"odinger equation which is
solved with a boundary condition (BC) at some $r=b$. This BC
is parametrized as an analytic function of energy for the
various partial waves. The BC parameters, representing
short-range physics, and the free parameters in the long-range
forces ({\it e.g.} the pion-nucleon coupling constant) are
determined from a fit to the data. In the ``standard'' Nijmegen
PWA's of Refs.~\cite{Ber90,Sto93} the boundary is put at $b=1.4$
fm, and the long-range strong potential outside of 1.4 fm is
taken as the OPE potential supplemented by the non-OPE forces
of the Nijmegen soft-core potential Nijm78~\cite{Nag78}. These
heavy-boson exchanges were included because OPE alone did not allow
for an optimal description of the data. In this standard $pp$ PWA,
we obtain with 19 BC parameters $\chi^2_{\rm min}=1968.7$ and
$f^2_{pp\pi^0}=0.0756(4)$, where the error is statistical, on
the Nijmegen 1998 $pp$ database below 350 MeV, in which 1951
$pp$ scattering data are included~\cite{WWW98}. This result
will serve here as a benchmark.

Let us demonstrate our method first with some parts of the
electromagnetic interaction. When one omits in the standard 1998
$pp$ PWA the magnetic-moment interaction, both from the potential
and in constructing the scattering amplitude, the $\chi^2_{\rm min}$
increases by 390.0 to $\chi^2_{\rm min}=2358.7$. This is
therefore a 19.7 standard deviation (s.d.) effect. Omitting
vacuum polarization leads to $\chi^2_{\rm min}=2181.3$, {\it i.e.},
a rise in $\chi^2_{\rm min}$ of 212.6, which corresponds to
14.6 s.d. These numbers demonstrate that one can use this method
of energy-dependent PWA to show the presence and the importance of
these specific well-known parts of the long-range $pp$ interaction.

A very important part of the energy dependence of the $N\!N$
phase shifts comes from OPE. In the Nijmegen energy-dependent
PWA's the different pion-nucleon coupling constants could be
determined accurately and reliably~\cite{Tim93,Ber87,Swa97}.
In Ref.~\cite{Swa97}, we recommended for the charge-independent
coupling constant the value $f_{N\!N\pi}^2 = 0.0750(9)$,
where the error includes statistical as well as systematic
effects. As a systematic check, the masses of the exchanged
pions were determined, with excellent results:
   $m_{\pi^0} = 135.6(1.0)$ MeV and
   $m_{\pi^+} = 139.6(1.3)$ MeV.
In this way, the presence of OPE in the $N\!N$ force was shown
with an enormous statistical significance. A more subtle effect
is the energy dependence of the OPE potential due to the
minimal-relativity factor $M/E$, where $M$ is the proton mass
and $E$ the proton center-of-mass energy. Omitting this factor
from the OPE potential results in $\chi^2_{\rm min}=1977.2$.
This is a rise of 8.5 in $\chi^2_{\rm min}$, or an almost 3 s.d.
effect. Recently, also the electromagnetic corrections to the
OPE potential in $np$ scattering were investigated~\cite{Kol98}.

The starting point to derive the OPE and TPE potentials
is the effective chiral Lagrangian, the leading-order
of which is the nonlinear Weinberg model~\cite{Wei66},
\begin{equation}
   {\mathscr L}^{(0)}
   = -\overline{N}\left[\,\gamma_\mu{\mathscr D}^\mu + M
   + g_Ai\gamma_5\gamma_\mu\,\vec{\tau}\cdot\vec{D}^\mu \,\right]N
   \ , \label{eq:PVWT}
\end{equation}
with the chiral-covariant derivatives~\cite{Wei96}
\begin{eqnarray}
  \vec{D}^\mu & = & D^{-1}\partial^\mu\vec{\pi}/F_\pi
  \ , \nonumber \\
  {\mathscr D}^\mu N & = & \left( \partial^\mu + \frac{i}{F_\pi}
  c_0\,\vec{\tau}\cdot\vec{\pi}\!\times\!\vec{D}^\mu \right) N
  \ . \label{eq:covder}
\end{eqnarray}
Here, $D=1+\vec{\pi}^2/F_\pi^2$, $g_A=1.2573$ is the
Gamow-Teller coupling, and $F_\pi=185$ MeV is the pion decay
constant; chiral symmetry fixes $c_0\equiv1$. Eq.~(\ref{eq:PVWT})
implies that the planar- and crossed-box TPE diagrams should be
calculated with the pseudovector (PV) $N\!N\pi$ Lagrangian. We
use the physical $N\!N\pi$ coupling constant $f$, {\it i.e.},
we trade in the Goldberger-Treiman value $g_A/F_\pi$ for
$\sqrt{4\pi}f/m_s$;
the scaling mass $m_s$ serves to make $f$ dimensionless and is
conventionally chosen to be numerically equal to the charged-pion
mass, $m_s\equiv m_{\pi^+}$. In addition to the PV $N\!N\pi$
interaction, Eq.~(\ref{eq:PVWT}) contains the Weinberg-Tomozawa
(WT) $N\!N2\pi$ seagull interaction~\cite{Tom66}, resulting in
triangle and football TPE diagrams.

In order to derive the TPE potential in subleading order,
three more $N\!N2\pi$ interactions are required~\cite{Ord92},
{\it viz}.
\begin{eqnarray}
  {\mathscr L}^{(1)}
  & = & -\overline{N}\left[\:
  8 c_1 D^{-1}m_\pi^2\vec{\pi}^2/F_\pi^2 +
  4 c_3\,\vec{D}_\mu\!\cdot\!\vec{D}^\mu \right. \nonumber \\
  &   & \left. + \,
  2 c_4\,\sigma_{\mu\nu}\,\vec{\tau}\cdot
  \vec{D}^\mu\!\times\!\vec{D}^\nu \:\right]N \ , \label{eq:c134}
\end{eqnarray}
leading to additional triangle diagrams. The values of the chiral
parameters (``low-energy constants'') $c_i$ ($i=1,3,4$) of order
$(1/M)$ are not fixed by chiral symmetry; the $c_i$'s
represent ``integrated-out'' hadrons, such as the heavier
mesons like the $\varepsilon$ and $\varrho$, and the $N$- and
$\Delta$-isobars. The definition Eq.~(\ref{eq:c134}) of these
$c_i$'s~\cite{Tim98} agrees with the convention used in
heavy-baryon $\chi$PT~\cite{Ber92,Eck96}; an additional
$c_2$-term does not contribute to the $N\!N$ force in this
order. The $c_1$-term violates chiral symmetry explicitly. A
systematic expansion of Eqs.~(\ref{eq:PVWT}) and (\ref{eq:c134})
to order $(1/M)$ gives the relevant part of the
chiral Lagrangian~\cite{Fri98}.

The OPE and TPE potentials derived from this Lagrangian
contain central, spin-spin, tensor, and spin-orbit terms,
{\it viz}.
\begin{equation}
  V = V_C + V_S\,\bfsig_1\cdot\bfsig_2 + V_T\,S_{12}
    + V_{SO}\,\boldsymbol{L}\cdot\boldsymbol{S} \ ,
\end{equation}
where $S_{12}=3\,\bfsig_1\cdot\bfr\,\bfsig_2\cdot\bfr
-\bfsig_1\cdot\bfsig_2$. With $i=C,S,T,SO$, $\xi=m_\pi/m_s$,
and $x=m_\pi r$, we can write
\begin{equation}
  V_i(r) = f^{2n}\,\xi^{2n} \left(M/E\right) \left[ v_i(x)
    + w_i(x)\vec{\tau}_1\!\cdot\!\vec{\tau}_2 \right] m_\pi \ ,
  \label{eq:V}
\end{equation}
where $n=1$ for OPE and $n=2$ for TPE.

The long-range OPE potential contains an isovector
spin-spin part $w_S$ and an isovector tensor part $w_T$,
\begin{eqnarray}
   w_S(x) & = & e^{-x}/3x \ , \nonumber \\
   w_T(x) & = & \left(1+x+x^2/3\right) e^{-x}/x^3 \ .
\end{eqnarray}
For the $pp$ case, the neutral-pion mass $m_{\pi^0}$ is used
in OPE. The coupling $f_p^2=f^2_{pp\pi^0}$ is a free parameter.

For TPE, the dimensionless isoscalar functions
$v_i$ are written as the sum of the leading-order terms
$v_{i,1}$ and the subleading-order terms $v_{i,2}$,
\begin{equation}
   v_i(x) = \left(2/\pi\right)\, v_{i,1}(x) +
            \left(m_\pi/M\right)\, v_{i,2}(x) \ ,
\end{equation}
and similarly for the isovector functions $w_i$. In the TPE
potential, we use the average pion mass $m_\pi=138.04$ MeV
and the fixed charge-independent coupling constant is $f^2 =
f^2_{N\!N\pi} = 0.0750$. Care must be taken to obtain the
appropriate form for the use of Eq.~(\ref{eq:V}) in the
relativistic Schr\"odinger equation. Other forms of the OPE
potential or other two-body equations will, in general, give
different TPE potentials~\cite{Rij91,Fri94}.

The leading-order static
potential TPE(l.o.) contains isoscalar spin-spin and tensor
terms, $v_{S,1}$ and $v_{T,1}$ respectively, and an isovector
central component $w_{C,1}$. The long-range parts are
\begin{eqnarray}
  v_{S,1}(x) & = & 12 K_0(2x)/x^3 +
             (12+8x^2)K_1(2x)/x^4 \ , \nonumber \\
  v_{T,1}(x) & = & -12 K_0(2x)/x^3 -
             (15+4x^2)K_1(2x)/x^4 \ , \label{eq:TMO} \\
  w_{C,1}(x) & = & \left( \tilde{c}_0^2+10\tilde{c}_0-23
             - 4x^2 \right) K_0(2x)/x^3 + \nonumber \\
             &   & \left( \tilde{c}_0^2+10\tilde{c}_0-23 +
             (4\tilde{c}_0-12)x^2 \right) K_1(2x)/x^4
 \nonumber \ ,
\end{eqnarray}
where the modified Bessel functions have asymptotic behavior
$K_n(2x)\sim\sqrt{\pi/4x}\,e^{-2x}$. This TPE(l.o.) is the
``TMO'' potential~\cite{Tak52}, supplemented by the diagrams
with the WT seagulls~\cite{Fri94,Epe98}. In the WT terms we
extracted, for ease of presentation, an overall factor $f^4$,
cf. Eq.~(\ref{eq:V}), and defined $\tilde{c}_0=c_0/\tilde{g}_A^2$,
where $\tilde{g}_A=F_\pi\sqrt{4\pi}f/m_s$.

The subleading-order potential TPE(s.o.) contains nonstatic
terms from Eq.~(\ref{eq:PVWT}) and the leading-order terms
from Eq.~(\ref{eq:c134}). The long-range parts read
\begin{equation}
   v_{i,2}(x) = \textstyle{\sum_{p=1}^6}\,a_p\,e^{-2x}/x^p \ ,
 \label{eq:coef}
\end{equation}
and similarly for $w_{i,2}$, with the coefficients $a_p$ as
collected in Table~\ref{tab:TPE}. Also here a factor $f^4$ was
extracted and the result was rewritten in terms of $\tilde{c}_0$
and $\tilde{c}_i=c_iM/\tilde{g}_A^2$. Our results for
TPE(s.o.) agree with Ref.~\cite{Kai97}.

Remarkably, a large part of the correct TPE potential
was already obtained by Sugawara and Okubo~\cite{Sug60}
in ``pre-chiral days,'' by using PV coupling and
two phenomenological $N\!N2\pi$ interactions:
the WT term of Eq.~(\ref{eq:PVWT}) and the $c_1$-part of
Eq.~(\ref{eq:c134}). They also pointed out that PV coupling
gives a rather strong attractive isoscalar spin-orbit force
in subleading order. However, the important additionally
required chiral $c_3$- and $c_4$-terms were missing; these
were for the $N\!N$ case first given in Ref.~\cite{Ord92}.

We now come to the results of the TPE studies, in which
we again use the 1998 database below 350 MeV, with 1951
data~\cite{WWW98}. The main results of the
various PWA's are summarized in Table~\ref{tab:PWA}. We start
conservatively with the boundary at $b=1.8$ fm, since beyond
1.8 fm only OPE and TPE are expected to contribute significantly.
When only OPE is included as strong force, $\chi^2_{\rm min}=1956.6$
is reached at the cost of 29 BC parameters. We want to investigate
if the fit can be even further improved when TPE is added. When
only the TPE(l.o.) potential of Eq.~(\ref{eq:TMO}) is used, we
obtain $\chi^2_{\rm min}=1965.9$ with $26$ BC parameters. But
we can do better. The complete TPE potential, $\chi$TPE =
TPE(l.o.) + TPE(s.o.), contains
three {\it a priori} unknown constants: the
chiral parameters $c_i$ ($i=1,3,4$) from Eq.~(\ref{eq:c134}).
In the fits we obtain $c_1=-4.4(3.4)$/GeV. The values of $c_1$
and $c_3$, appearing both only in the isoscalar central potential,
cf. Table~\ref{tab:TPE}, are strongly correlated. The correlations
between the parameters can be summarized concisely by:
\begin{eqnarray*}
  c_3 & = & \left[ -5.08 - 0.62 (c_1+0.76)
                  + 40 (f_p^2-0.0755) \right]/{\rm GeV} , \\
  c_4 & = & \left[ +4.70 + 0.01 (c_1+0.76)
                  + 250 (f_p^2-0.0755) \right]/{\rm GeV} .
\end{eqnarray*}
In order to determine reliable values for $c_3$ and $c_4$,
we use the theoretical estimate~\cite{Ber92} for $c_1$
obtained from the scalar form factor $\sigma(t)$ of the
proton~\cite{Gas91} at $t=0$, {\it viz}.
\begin{equation}
  c_1 = -\left[ \sigma(0)/4m_\pi^2 +
        9f^2\xi^2/16m_\pi \right] \ ; \label{eq:sig0}
\end{equation}
$\sigma(0)$ is the pion-nucleon sigma term, the value of
which is uncertain. We take here the plausible ``low'' value
$\sigma(0)=35(5)$ MeV~\cite{Gas81}, which is supported by
the recent $\pi N$ PWA of Ref.~\cite{Tim97}. This gives
\begin{equation}
  c_1 = -[ 0.46(7) + 0.30 ]/{\rm GeV} = -0.76(7)/{\rm GeV} \ ;
  \label{eq:c1}
\end{equation}
the error here is theoretical. Our determination of $c_1$ is
consistent with this value. Fixing $c_1=-0.76$/GeV, we find,
with 22 BC parameters, $\chi^2_{\rm min}=1937.8$ and
$f_p^2=0.0755(7)$; the resulting values for $c_3$ and $c_4$ are
\begin{equation}
   c_3 = -5.08(28)/{\rm GeV} \ , \;\;\;
   c_4 = +4.70(70)/{\rm GeV} \ , \label{eq:c34}
\end{equation}
where the errors are statistical. The improvement over
only OPE is reflected, even beyond 1.8 fm, in the 18.8 lower
$\chi^2_{\rm min}$ and in the 7 fewer BC parameters required.

The result found for $f_p^2$ is in very good
agreement with the value 0.0756(4) determined in the standard
1998 $pp$ PWA. Our values for the $c_i$'s can be compared to
the determination from the $\pi N$ scattering amplitudes in
Ref.~\cite{Ber97}. Here, $c_1=-0.93(9)$/GeV was obtained using
Eq.~(\ref{eq:sig0}), but with $\sigma(0)=45(8)$ MeV, along with
$c_3=-5.29(25)/$GeV and $c_4=+3.63(10)$/GeV. In view of the
uncertainties in the $\pi N$ amplitudes~\cite{Tim97}, the
good agreement is a significant success. It underlines, for
the first time quantitatively, that the long-range $N\!N$ and
the low-energy $\pi N$ interactions are governed by the same
chiral Lagrangian.

In previous studies of the OPE potential,
a good systematic check has been the determination of the
masses of the exchanged pions. In order to check explicitly
that we are now actually looking at the TPE interaction, we
determine the range. This is done by adding the pion mass
$m_\pi$ in the potential $\chi$TPE as another free parameter.
We first fix the pion coupling in OPE at $f_p^2=0.0755$ and the
$c_i$'s to their central values given in Eqs.~(\ref{eq:c1}) and
(\ref{eq:c34}). Then we fit an overall scale factor $\lambda$
for the potential $\chi$TPE, the pion mass $m_\pi$, and
the BC parameters. The results are: $\lambda=0.82(16)$ and
$m_\pi=125(10)$ MeV. Alternatively, we fix $c_1$
and fit $m_\pi$ together with $f_p^2$, $c_3$, $c_4$, and
the BC parameters. This results in $m_\pi=128(9)$ MeV, again
in good agreement with the average pion mass $m_\pi=138.04$ MeV.
The very good $\chi^2_{\rm min}$ obtained, the good values for
the $c_i$'s, and this correct pion mass constitute convincing
proof for the presence of chiral TPE loops in the long-range
$pp$ interaction.

In order to investigate further the importance of $\chi$TPE,
we move the boundary inwards to $b=1.4$ fm. When only OPE is
used as long-range force, it is possible to achieve
a reasonable fit: at the cost of 31 BC parameters
$\chi^2_{\rm min}=2026.2$ is reached. We then add to OPE
the potential TPE(l.o.). With
28 BC parameters, $\chi^2_{\rm min}=1984.7$ is obtained.
Compared to only OPE, this corresponds to a drop in
$\chi^2_{\rm min}$ of $41.5$ with 3 fewer parameters,
a significant improvement. However, the fit is still
not optimal~\cite{Ren94}. We next add also the potential
TPE(s.o.). With fixed $c_1=-0.76$/GeV,
this gives with 23 BC parameters $\chi^2_{\rm min}=1934.5$,
$c_3=-4.99(21)$/GeV, and $c_4=+5.62(59)$/GeV. This shows that
OPE together with $\chi$TPE gives a very good $N\!N$ force at
least as far inwards as 1.4 fm.

In conclusion, we have, for the first time, incorporated and
studied chiral TPE in an energy-dependent PWA of the $pp$ scattering
data. The main result of this Letter is that we have shown the
presence of chiral TPE loops in the long-range $pp$ interaction.
A significant improvement over using just OPE is seen. With OPE
and $\chi$TPE,
an excellent fit to the database becomes possible, even somewhat
better than the standard 1998 $pp$ PWA. The chiral parameters
agree with those found in $\pi N$ scattering. Especially important
in obtaining the very good fit is the isoscalar central attraction
from the $c_3$-term, partly a ``chiral van-der-Waals force'' due
to the axial polarizability of the nucleon~\cite{Tar78}.
In all, our results provide a big success for chiral symmetry.
A novel class of PWA has been established, with such a
theoretically well-founded and model-independent chiral TPE
potential included in all partial waves.

We thank U. van Kolck and Th.A. Rijken for helpful discussions.
The research of R.T.\ was made possible by a fellowship of the
Royal Netherlands Academy of Arts and Sciences; that of J.L.F.\
was performed under the auspices of the United States Department
of Energy. M.R.\ acknowledges the financial support of the
Australian Research Council and the hospitality of the KVI.

\clearpage
\widetext
\begin{table}
\caption{Coefficients of the subleading-order potential TPE(s.o.)
         of Eq.~(\ref{eq:coef}), for the central, spin-spin,
         tensor, and spin-orbit terms, both isoscalar and
         isovector. We defined $\tilde{c}_0=c_0/\tilde{g}_A^2$,
         $\tilde{c}_i=c_iM/\tilde{g}_A^2$ for $i=1,3,4$,
         and $\tilde{c}_{04}=\tilde{c}_0+4\tilde{c}_4$.}
\begin{tabular}{cl|rrrrrr}
     &      & \multicolumn{6}{c}{$a_p$} \\
     & $i$  & $p=1$ & $p=2$ & $p=3$ & $p=4$ & $p=5$ & $p=6$
 \\ \hline
 $v$ & $C$  & $3/4$
            &     $9+48\tilde{c}_1+24\tilde{c}_3$
            &    $27+96\tilde{c}_1+96\tilde{c}_3$
            & $99/2+48\tilde{c}_1+240\tilde{c}_3$
            &                 $54+288\tilde{c}_3$
            &                 $27+144\tilde{c}_3$ \\
     & $S$  & &  $-3$ &   $-9$ & $-33/2$ & $-18$ &  $-9$ \\
     & $T$  & & $3/2$ & $27/4$ &    $15$ &  $18$ &   $9$ \\
     & $SO$ & &       &  $-12$ &   $-36$ & $-48$ & $-24$ \\
 $w$ & $C$  & $3/2$
            &   $4-2\tilde{c}_0$
            &  $14-8\tilde{c}_0$
            & $31-20\tilde{c}_0$
            & $36-24\tilde{c}_0$
            & $18-12\tilde{c}_0$ \\
     & $S$  & & $-2/3$
            &  $-14/3+8\tilde{c}_{04}/3$
            & $-31/3+20\tilde{c}_{04}/3$
            &        $-12+8\tilde{c}_{04}$
            &         $-6+4\tilde{c}_{04}$ \\
     & $T$  & & $1/3$
            &  $17/6-4\tilde{c}_{04}/3$
            & $26/3-16\tilde{c}_{04}/3$
            &        $12-8\tilde{c}_{04}$
            &         $6-4\tilde{c}_{04}$ \\
     & $SO$ & & &
            &   $8-8\tilde{c}_0$
            & $16-16\tilde{c}_0$
            &   $8-8\tilde{c}_0$ \\
\end{tabular}
\label{tab:TPE}
\end{table}

\narrowtext
\begin{table}
\caption{Results for the PWA's with different long-range
         interactions. $\#$BC is the number of BC parameters.}
\begin{tabular}{l|cccc}
                 & \multicolumn{2}{c}{$b=1.4$ fm}
                 & \multicolumn{2}{c}{$b=1.8$ fm} \\
                 & $\#$BC & $\chi^2_{\rm min}$
                 & $\#$BC & $\chi^2_{\rm min}$  \\ \hline
 Nijm78          &  19  & 1968.7 &  $-$&  $-$   \\
 OPE             &  31  & 2026.2 &  29 & 1956.6 \\
 OPE + TPE(l.o.) &  28  & 1984.7 &  26 & 1965.9 \\
 OPE + $\chi$TPE &  23  & 1934.5 &  22 & 1937.8 \\
\end{tabular}
\label{tab:PWA}
\end{table}


\begin{references}
\bibitem{Tim98} In Ref.~\cite{Ord92}, the chiral parameters
                are named $B_i$ ($i=1,2,3$), where
                $B_1=4c_3$, $B_2=4c_4+1/M$, and $B_3=8c_1$.
\end{references}
\end{document}